\documentclass[sigconf]{acmart}

\usepackage{booktabs} 
\usepackage{graphicx}
\usepackage[show]{chato-notes}
\usepackage{amsmath,amssymb}
\usepackage{epsfig}
\usepackage{algorithm}
\usepackage[noend]{algorithmic}
\usepackage{url}
\usepackage{hyperref}
\usepackage{xspace}
\usepackage{balance}
\usepackage{booktabs}

\makeatletter
\newsavebox\myboxA
\newsavebox\myboxB
\newlength\mylenA

\newcommand*\xoverline[2][0.75]{%
    \sbox{\myboxA}{$\m@th#2$}%
    \setbox\myboxB\null
    \ht\myboxB=\ht\myboxA%
    \dp\myboxB=\dp\myboxA%
    \wd\myboxB=#1\wd\myboxA
    \sbox\myboxB{$\m@th\overline{\copy\myboxB}$}
    \setlength\mylenA{\the\wd\myboxA}
    \addtolength\mylenA{-\the\wd\myboxB}%
    \ifdim\wd\myboxB<\wd\myboxA%
       \rlap{\hskip 0.5\mylenA\usebox\myboxB}{\usebox\myboxA}%
    \else
        \hskip -0.5\mylenA\rlap{\usebox\myboxA}{\hskip 0.5\mylenA\usebox\myboxB}%
    \fi}
\makeatother

\hypersetup{
    unicode=false,          
    pdftoolbar=true,        
    pdfmenubar=true,        
    pdffitwindow=false,     
    pdfstartview={FitH},    
    pdftitle={My title},    
    pdfauthor={Author},     
    pdfsubject={Subject},   
    pdfcreator={Creator},   
    pdfproducer={Producer}, 
    pdfnewwindow=true,      
    colorlinks=true,       
    linkcolor=black,          
    citecolor=black,        
    filecolor=black,      
    urlcolor=black           
}

\newcommand{\spara}[1]{\smallskip\noindent{\bf #1}}

\newtheorem{mydefinition}{Definition}

\newcommand{\NP}{$\mathbf{NP}$\xspace}
\newcommand{\NPhard}{$\mathbf{NP}$-hard\xspace}

\newcommand{\spancore}{span-core\xspace}

\newcommand{\spancores}{\spancore{s}\xspace}

\newcommand{\corenessvec}{coreness vector}

\renewcommand{\vec}[1]{\mathbf{#1}}

\newcommand{\squishlist}{
 \begin{list}{$\bullet$}
  {  \setlength{\itemsep}{0pt}
     \setlength{\parsep}{3pt}
     \setlength{\topsep}{3pt}
     \setlength{\partopsep}{0pt}
     \setlength{\leftmargin}{2em}
     \setlength{\labelwidth}{1.5em}
     \setlength{\labelsep}{0.5em}
} }
\newcommand{\squishlisttight}{
 \begin{list}{$\bullet$}
  { \setlength{\itemsep}{0pt}
    \setlength{\parsep}{0pt}
    \setlength{\topsep}{0pt}
    \setlength{\partopsep}{0pt}
    \setlength{\leftmargin}{2em}
    \setlength{\labelwidth}{1.5em}
    \setlength{\labelsep}{0.5em}
} }

\newcommand{\squishdesc}{
 \begin{list}{}
  {  \setlength{\itemsep}{0pt}
     \setlength{\parsep}{3pt}
     \setlength{\topsep}{3pt}
     \setlength{\partopsep}{0pt}
     \setlength{\leftmargin}{1em}
     \setlength{\labelwidth}{1.5em}
     \setlength{\labelsep}{0.5em}
} }

\newcommand{\squishend}{
  \end{list}
}

\copyrightyear{2019}
\acmYear{2019} 
\setcopyright{iw3c2w3}
\acmConference[WWW '19 Companion]{Companion Proceedings of the 2019 World Wide Web Conference}{May 13--17, 2019}{San Francisco, CA, USA}
\acmBooktitle{Companion Proceedings of the 2019 World Wide Web Conference (WWW '19 Companion), May 13--17, 2019, San Francisco, CA, USA}
\acmPrice{}
\acmDOI{10.1145/3308560.3314190}
\acmISBN{978-1-4503-6675-5/19/05}

\begin{document}

\title{Cores and Other Dense Structures in Complex Networks}

\author{Edoardo Galimberti}
\affiliation{%
\institution{University of Turin \& ISI Foundation, Italy}
}
\email{edoardo.galimberti@unito.it}

\begin{abstract}
Complex networks are a powerful paradigm to model complex systems. 
Specific network models, e.g., \emph{multilayer networks}, \emph{temporal networks}, and \emph{signed networks}, enrich the standard network representation with additional information to better capture real-world phenomena.
Despite the keen interest in a variety of problems, algorithms, and analysis methods for these types of network, the problem of extracting cores and dense structures still has unexplored facets.

In this work, we present advancements to the state of the art by the introduction of novel definitions and algorithms for the extraction of dense structures from complex networks, mainly cores.
At first, we define \emph{core decomposition in multilayer networks} together with a series of applications built on top of it, i.e., the extraction of \emph{maximal multilayer cores} only, \emph{densest subgraph} in multilayer networks, the speed-up of the extraction of \emph{frequent cross-graph quasi-cliques}, and the generalization of \emph{community search} to the multilayer setting.
Then, we introduce the concept of \emph{core decomposition in temporal networks}; also in this case, we are interested in the extraction of \emph{maximal temporal cores} only.
Finally, in the context of discovering polarization in large-scale online data, we study the problem of identifying \emph{polarized communities in signed networks}.

The proposed methodologies are evaluated on a large variety of real-world networks against na\"{\i}ve approaches, non-trivial baselines, and competing methods.
In all cases, they show effectiveness, efficiency, and scalability.
Moreover, we showcase the usefulness of our definitions in concrete applications and case studies, i.e., the temporal analysis of contact networks, and the identification of polarization in debate networks.
\end{abstract}



\maketitle
\sloppy

\section{Problem}
\label{sec:problem}
Complex networks model a large variety of systems of high technological and intellectual importance, such as the Internet, the World Wide Web, biological and chemical systems, financial, social, transportation, contact, and communication networks.
The need to obtain better understanding of such complex systems summed with their inherent complexity are factors that explain the increasing interest in enhancing complex-networks analytics tools and algorithms.
Many network models enrich the standard network representation with additional information that can lead to capturing novel and interesting structural properties of real-world phenomena.
For example, \emph{multilayer networks} model complex systems where various types of relations might occur among the same set of entities;
\emph{temporal networks} are representations of entities, their relations, and how these relations are established/broken along time;
and, \emph{signed networks} represent networked data where edge annotations express whether each edge interaction is friendly (positive) or antagonistic (negative).

Extracting dense structures from large graphs has emerged as a key graph-mining primitive in a variety of scenarios~\cite{aggarwal}, ranging from web mining, to biology, and finance.
For instance, dense patterns can help in studying contact networks among individuals to quantify the transmission opportunities of respiratory infections.
Anomalously dense patterns among entities in a co-occurrence network have also been used to identify, in real-time, events and buzzing stories.
Peculiar dense structures are often associated to polarization and unbalance in online debate networks.
Although the literature about complex networks has recently grown extremely fast, the problem of identifying cores and other dense structures in specific types of networks has, surprisingly, still unexplored facets.

Among the many definitions of dense structures, \emph{core decomposition} plays a central role.
The $k$-\emph{core} of a graph is defined as a maximal subgraph in which every vertex is connected to at least $k$ other vertices within such subgraph.
The set of all $k$-cores of a graph $G$ forms the \emph{core decomposition} of $G$~\cite{Seidman1983k-cores}.
The importance of core decomposition relies in the fact that it can be computed in linear time~\cite{batagelj2011fast}, and can be used to speed-up/approximate dense-subgraph extraction according to various other definitions.
In addition, it has been recognized as an important tool to analyze and visualize complex networks in several domains.

Orthogonal approaches focus on the identification of a single dense portion of a complex network.
Several notions of \emph{density} exist in the literature, each of which leads to a different version of the problem.
While most variants are \NPhard, or even inapproximable, extracting dense subgraphs according to the \emph{average-degree density} (i.e., two times the number of edges divided by the number of vertices) is solvable in polynomial time~\cite{Goldberg84}.
As a result, such a density has attracted most of the research in the field, so that the subgraph maximizing the average-degree density is commonly referred to as the \emph{densest subgraph}.

A different bulk of literature deals with methods that look for finding many communities, i.e., dense structures, while partitioning the whole complex network.
Among the many approaches, of particular interest is the one provided by the {\em correlation clustering} framework~\cite{bansal2004correlation} for signed networks, which asks to partition the vertices into communities so as to maximize (minimize) the number of edges that ``agree'' (``disagree'') with the partitioning, i.e., the number of positive (negative) edges within clusters plus the number of negative (positive) edges across clusters.

In this work, we propose a set of novel definitions of dense structures in complex networks with particular attention on the algorithmic aspects.
Specifically, we study ($i$) the problem of \emph{core decomposition in multilayer networks} together with a series of applications built on top of it; ($ii$) the notion of \emph{core decomposition in temporal network}; and, ($iii$) the problem of identifying \emph{polarized communities in signed networks}.
The original work about multilayer core decomposition is enclosed in~\cite{GalimbertiBG17}; the enriched version of the same work has been submitted to the TKDD Journal~\cite{tkddmulti}.
The work about temporal core decomposition is included in~\cite{galimberti2018mining}, while the one regarding polarized communities is currently under review.

\section{State of the art}
\label{sec:state}
\spara{Core decomposition.}
Given a simple graph $G=(V,E)$, let $d(S,u)$ denote the degree of vertex $u \in V$ in the subgraph induced by vertex set $S \subseteq V$, i.e., $d(S,u) = |\{v \in S \mid (u,v) \in E \}|$.

\vspace{-1mm}
\begin{mydefinition}[Core Decomposition]\label{def:kcores}
The $k$\emph{-core} (or core of order $k$) of $G$ is a maximal set of vertices $C_k \subseteq V$ such that $\forall u \in C_k: d(C_k,u) \geq k$.
The set of all $k$-cores $V = C_0 \supseteq C_1 \supseteq \cdots \supseteq C_{k^*}$ ($k^* = \arg\max_{k} C_k \neq \emptyset$) is the \emph{core decomposition} of $G$.
\end{mydefinition}
\vspace{-1mm}

Core decomposition has been recognized as an important tool to analyze and visualize complex networks~\cite{DBLP:conf/gd/BatageljMZ99,Alvarez-HamelinDBV05} in several domains, e.g., bioinformatics, software engineering, and social networks.
It has been studied under various settings, such as distributed, streaming/maintenance, and disk-based, and for various types of graphs, such as uncertain, directed, and weighted.
The maximal core, i.e., the core of highest order, has been shown to have an important role in determining the propagation patterns of epidemic processes on networks~\cite{Kitsak2010}.

Azimi-Tafreshi~{\em et~al.}~\cite{MultiplexCores} study the core-percolation problem in multilayer networks from a physics standpoint without providing any algorithm.
They characterize cores on 2-layer Erd\H{o}s-R\'{e}nyi and 2-layer scale-free networks, then they analyze real-world (2-layer) air-transportation networks.
A type of core decomposition for temporal networks has been proposed by Wu~{\em et~al.}~\cite{wu2015core}, who define the $(k,h)$-core  as the largest subgraph in which every vertex has at least $k$ neighbors and at least $h$ temporal connections with each of them.
Finally, Zhang~\emph{et~al.}~\cite{zhang2017engagement} study the problem of enumerating all maximal cores of a (non-temporal) variant of core decomposition, that turns out to be $\mathbf{NP}$-hard.

\spara{Densest subgraph.}
Goldberg~\cite{Goldberg84} provides an exact solution to the densest-subgraph problem based on iteratively solving ad-hoc-defined minimum-cut problem instances.
Asahiro~\emph{et~al.}~\cite{AITT00} and Charikar~\cite{Char00} introduce a more efficient (linear-time) $\frac{1}{2}$-approximation algorithm that resembles the one for core decomposition.
In fact, it can be proved that the maximal core of a graph is itself a $\frac{1}{2}$-approximation of the densest subgraph.

In the classic definition of densest subgraph there is no size restriction of the output.
Variants of the problem with size constraints turn out to be \NPhard.
A number of works depart from the classic average-degree maximization problem and focus on extracting a subgraph maximizing other notions of density, e.g., quasi-clique-based density, or triangle density.
The densest-subgraph problem has also been studied in different settings, such as streaming/dynamic context, and top-$k$ fashion.

\spara{Dense structures in multilayer networks.}
Jethava~and~Beerenwinkel~\cite{jethava2015finding} define the densest common subgraph problem, i.e., find a subgraph maximizing the minimum average degree over all layers of the input graph, and devise a linear-programming formulation and a greedy heuristic algorithm for it.
A very recent work by Charikar~\emph{et~al.}~\cite{charikar2018finding} further focuses on the minimum-average and average-minimum formulations by providing several theoretical findings, including \NPhard{ness} and hardness of the approximation (for both minimum-average and average-minimum).
Jiang~\emph{et~al.}~\cite{jiang2009mining} focus on extracting frequent cross-graph quasi-cliques, i.e., subgraphs that are quasi-cliques in at least a fraction of layers equal to a certain minimum support and have size larger than a given threshold.
Interdonato~\emph{et~al.}~\cite{InterdonatoTISP17} are the first to study the problem of local community detection in multilayer networks, i.e., when a seed vertex is given and we want to reconstruct its community by having only a limited local view of the network.

\spara{Dense structures in temporal networks.}
Semertzidis~\emph{et~al.}~\cite{semertzidis2016best} introduce the problem of identifying a set of vertices that are densely connected in all or at least $k$ timestamps of a temporal network.
The notion of $\Delta$-clique has been proposed in~\cite{viard2016computing}, as a set of vertices in which each pair is in contact at least every $\Delta$ timestamps.
Complementary approaches study the problem of discovering dense temporal subgraphs whose edges occur in short time intervals considering the exact timestamp of the occurrences, and the problem of maintaining the densest subgraph in a dynamic setting.
A slightly different yet related body of literature focuses on frequent evolution patterns in temporal attributed graphs, link-formation rules in temporal networks, and the discovery of dynamic relationships and events, or of correlated activity patterns.
Bogdanov~\emph{et~al.}~\cite{bogdanov2011mining} study the problem of finding a subgraph that maximizes the sum of edge weights in a network whose topology remains fixed but edge weights evolve over time.

\spara{Balance theory.}
The concept of signed graph in graph theory appeared in a work by Harary, who was particularly interested in the notion of balance in graphs.
Cartwright and Harary generalized Heider's psychological theory of balance in triangles of sentiments to a psychological theory of balance in signed graphs.
Harary and Kabell develop a simple linear-time algorithm to test whether a given signed graph satisfies the balance property~\cite{harary1980simple}, namely all cycles in the graph contain an even number of negative edges.
A more recent line of work develops spectral properties of signed graphs, related to balance theory.
Hou~\emph{et~al.}~\cite{hou2003laplacian} prove that a connected signed graph is balanced if and only if the least Laplacian eigenvalue is~0.
Subsequently, Hou~\cite{hou2005bounds} also investigates the relationship between the least Laplacian eigenvalue and the unbalancedness of a signed graph.

\spara{Correlation clustering and applications of signed networks.}
In the original correlation-clustering problem formulation~\cite{bansal2004correlation}, the number of clusters is not given as input, instead it is part of the optimization.
More recent works study the correlation-clustering problem with additional constraints, e.g., a fixed number of clusters or bounds on the cluster sizes.
Coleman~\emph{et~al.}~\cite{coleman2008local} employ the correlation-clustering framework to search for exactly two communities partitioning the whole network.

Signed graphs have also been studied, in different contexts, in machine learning and data mining.
Leskovec~\emph{et~al.}~\cite{leskovec2010signed} study directed signed graphs and develop status theory, which complements balance theory, to reason about the importance of the vertices in such graphs.
Other lines of research include edge and vertex classification, link prediction, community detection, recommendation, and more.
A detailed survey on the topic is provided by Tang~\emph{et~al.}~\cite{tang2016survey}.

\vspace{-1mm}
\section{Proposed approach}
\vspace{-1mm}
\label{sec:approach}
\spara{Core decomposition in multilayer networks.}
A first step towards the identification of novel core/dense structures in complex networks is the study the problem of \emph{core decomposition in multilayer networks}.
Let $G = (V,E,L)$ be a multilayer graph, where $V$ is a set of vertices, $L$ is a set of layers, and $E \subseteq V \times V \times L$ is a set of edges.
Given an $|L|$-dimensional integer vector $\vec{k} = [k_{\ell}]_{\ell \in L}$, the \emph{multilayer} $\vec{k}$-\emph{core} of $G$ is  a maximal subgraph whose vertices have at least degree $k_{\ell}$ in such subgraph, for all layers $\ell \in L$.
Vector $\vec{k}$ is dubbed \emph{\corenessvec} of that core.
The set of all non-empty and distinct multilayer cores constitutes the \emph{multilayer core decomposition} of $G$.
This definition is studied in~\cite{MultiplexCores} from a physics standpoint without providing any algorithm.
To the best of our knowledge, no prior work has studied how to efficiently compute the complete core decomposition of multilayer networks.

As subsequent step in this direction, we present a series of applications built on top of multilayer core decomposition.
First we focus on the problem of extracting only the \emph{maximal multilayer cores}, i.e., cores that are not dominated by any other core in terms of coreness-vector indexes on all the layers.
Since maximal cores are orders of magnitude less than all the cores, it is interesting to develop algorithms that effectively exploit the maximality property and extract maximal cores directly, without first computing a complete decomposition. 
Then, we show how multilayer core decomposition finds application to the problem of \emph{densest-subgraph extraction} from multilayer networks~\cite{jethava2015finding,charikar2018finding}.
As a further application, we exploit multilayer core decomposition to speed-up the identification of \emph{frequent cross-graph quasi-cliques}~\cite{jiang2009mining}.
Finally, we show how multilayer core decomposition can be used to generalize the \emph{community-search} problem \cite{Sozio} to the multilayer setting.

\spara{Core decomposition in temporal networks.}
For temporal networks, we adopt as measure of density of a pattern the \emph{minimum degree} holding among the vertices of a subgraph during a temporal span. 
The problem of extracting all such patterns is tackled by introducing a notion of \emph{temporal core decomposition} in which each core is associated with its \emph{span}, i.e., an interval of contiguous timestamps, for which the coreness property holds.
We call such cores \emph{span-cores}.
In all prior works~\cite{wu2015core,GalimbertiBG17}, the sequentiality of connections is not taken into account since non-contiguous timestamps can support the same core, then cores cannot be assigned with a clear temporal collocation.

As the total number of time intervals is quadratic in the size of the temporal domain $T$ under analysis, also the total number of span-cores is, in the worst case, quadratic in $T$, which may be too large an output for human inspection.
In this regard, we shift our attention to the problem of finding only the \emph{maximal span-cores}, i.e., span-cores that are not dominated by any other span-core by both the coreness property and the span.
Zhang~\emph{et~al.}~\cite{zhang2017engagement} also introduce a notion of maximal cores, but theirs cannot be identified in polynomial time (their problem formulation is \NPhard) and are not associated with a span.

\spara{Polarized communities in signed networks.}
In the context of studying polarization in large-scale online data, one of the fundamental data-analysis tasks is the {\em detection of polarization}.
We then employ singed networks to study a fundamental problem abstraction for this task, in particular, the problem of \emph{discovering polarized communities}.

In this scenario of application, we propose the problem of finding two communities such that ($i$) within communities there are mostly positive edges while across communities there are mostly negative edges, and ($ii$) the communities are embedded within a large body of other network vertices, which are neutral with respect to the polarization structure.
Our hypothesis is that such two-community polarized structure captures accurately controversial discussions in real-world social-media environments.
To the best of our knowledge, our work is the first aiming at extracting two hidden polarized communities from signed networks.
Our problem formulation deviates from the bulk of the literature where methods typically look for finding many communities while partitioning the whole network.
The closest proposal to our problem statement is the work by Coleman~\emph{et~al.}~\cite{coleman2008local}, who employ the correlation-clustering framework and search for exactly two communities, but again they aim at partitioning the whole network.

\vspace{-1mm}
\section{Methodology}
\vspace{-1mm}
\label{sec:methodology}
\spara{Core decomposition in multilayer networks.}
A major challenge of computing the complete core decomposition of multilayer networks is that \emph{the number of multilayer cores can be exponential in the number of layers}.
In fact, unlike the single-layer case where cores are all nested into each other, no total order exists among multilayer cores.
Rather, they form a core lattice defining a relation of partial containment.
Within this view, our first contribution is to devise three algorithms that exploit effective pruning rules during the visit of the lattice, thus being much more efficient than a na\"{\i}ve counterpart that computes every core starting from the whole input graph.

Shifting the attention to the problem of computing all and only the maximal cores, a straightforward way of approaching this problem would be to first compute the complete core decomposition, and then filter out the non-maximal cores.
However, as the maximal cores are usually much less than the overall cores, it would be desirable to have a method that effectively exploits the maximality property and extracts the maximal ones directly, without computing a complete decomposition.
In this work we show that, by means of a clever core-lattice visiting strategy, we can prune huge portions of the search space, thus achieving higher efficiency than computing the whole decomposition.

As a major application of multilayer core decomposition, we focus on the problem of extracting the densest subgraph from a multilayer network.
We generalize the problem studied in~\cite{charikar2018finding,jethava2015finding} by introducing a formulation that accounts for a trade-off between high density and number of layers exhibiting such high density.
The importance of the two ingredients of the objective is governed by a parameter $\beta$.
At first, we prove that the problem is \NP-hard.
Then, we show that computing the multilayer core decomposition of the input graph and selecting the core maximizing the proposed multilayer density function achieves a $\frac{1}{2|L|^\beta}$-approximation for the general multilayer-densest-subgraph problem formulation, and a $\frac{1}{2}$-approximation for the all-layer specific variant studied in~\cite{jethava2015finding,charikar2018finding}.

As a further application of our multilayer core-decomposition tool, we show how it can be used as a profitable preprocessing step to speed-up the problem of extracting frequent cross-graph quasi-cliques defined in~\cite{jiang2009mining}.
Specifically, we prove that the search space of the frequent cross-graph quasi-cliques can be circumstantiated to a number of restricted areas of the input multilayer graph, corresponding to multilayer cores that comply with the quasi-clique conditions.
This allows for skipping the visit of unnecessary parts of the input graph, and, thus, speeding-up the whole process, no matter which specific algorithm is used.

Finally, we also provide a generalization of the community-search problem~\cite{Sozio} to the multilayer setting, and show how to exploit multilayer core decomposition to obtain optimal solutions to this problem.

\spara{Core decomposition in temporal networks.}
Given a temporal domain $T$, the number of time intervals included in $T$ is quadratic in the duration of $T$.
Thus, since \emph{a span-core can possibly exist in every time interval}, a $|T|^2$ number of standard core decompositions would in principle be required to compute all the span-cores.
As a contribution to the problem of computing all \spancores, we derive containment properties among span-cores, and exploit them to devise an algorithm that computes a span-core decomposition much more efficiently than the na\"{\i}ve approach.

Then, we turn our attention to the problem of finding only the maximal span-cores.
As in the multilayer case, a straightforward way of approaching the maximal-\spancore-mining problem is to filter out non-maximal \spancores during the execution of an algorithm for computing the whole \spancore decomposition.
Even in the temporal setting, we can derive a number of theoretical properties about the relationship among \spancores of different temporal intervals and, based on these findings, we show a more efficient way of exploiting the maximality property and extracting maximal \spancores directly, without computing a complete decomposition.

\spara{Polarized communities in signed networks.}
From a theoretical standpoint, we formalize the problem of identifying two polarized communities in a signed network as a \emph{discrete eigenvector} problem.
The objective chosen in our formalization is penalized with the size of the solution in order to induce solutions of smaller dimension.
Therefore, vertices are only added to one of the two communities if they contribute significantly to the objective.
Also, our problem does not enforce balance between the communities.
This can be beneficial if there exist polarized communities of significantly different size in the input network.

We prove that our problem formulation is \NP-hard exploiting a reduction to classic correlation clustering, and propose two intuitive approximation algorithm based on spectral theory.
One is deterministic with quality guarantee $n$, where $n$ is the number of vertices in the graph.
It works by simply discretizing the entries of the singular vector $\vec{v}$ of the adjacency matrix corresponding to the largest singular value.
The other one is randomized with quality guarantee $\sqrt{n}$:
it randomly includes each vertex to one of the two communities with probabilities determined by the entries of $\vec{v}$.
Algorithms running time is essentially the time required to compute the first eigenvector of the adjacency matrix of the input graph.
Finally, we propose tweaks to enhance the flexibility of our algorithms and produce a wider variety of results.

\vspace{-1mm}
\section{Results}
\vspace{-1mm}
\label{sec:results}
\spara{Core decomposition in multilayer networks.}
We evaluate our algorithms for mining all multilayer cores and maximal multilayer cores only against their na\"{\i}ve counterparts on real-world multilayer networks in terms of runtime, memory consumption, and search-space exploration.
In all cases, the proposed methods result to be more efficient than the baselines, up to four orders of magnitude.
We also characterize all multilayer cores and maximal multilayer cores only.
As expected, the number of maximal cores is extremely lower: it ranges from the 0.3\% to the 22\% of the number of all cores.

We also verify the goodness of the proposed applications of multilayer core decomposition.
Experimental evidences about our definition of multilayer densest subgraph show that the correct tuning of the parameter $\beta$ is a key factor in obtaining interesting solutions in real-world networks.
Our pruning approach to the extraction of frequent cross-graph quasi-cliques results to be extremely effective (it prunes up to the 99\% of the input multilayer graph), being orders of magnitude fastest than the original procedure.
Finally, we evaluate the performance of our algorithms for multilayer community search varying the number of query vertices: in all cases they are able to return a solution in a reasonable amount of time, considerably lower than computing the whole decomposition.

\spara{Core decomposition in temporal networks.}
We carry out a comprehensive experimentation on several real-world temporal networks, with millions of vertices, tens of millions of edges, and hundreds of timestamps, to attest efficiency and scalability of our methods.
Our algorithms are able to reach up to two orders of magnitude of speed-up with respect to the na\"{\i}ve counterparts.

In order to showcase the usefulness of the proposed definitions, we apply (maximal) span-cores in the analysis of face-to-face contact networks gathered in primary and high schools.
Span-cores yield a straightforward temporal analysis of social activities of groups of people within a school day: they are able to identify class hours and breaks, and interesting mixing patterns of students with respect to gender and class.
We also analyze the duration of interactions of social groups in schools by studying the distribution of the size of the spans of the maximal span-cores.
This analysis confirms a robust statistical behavior and has similar results with respect to simpler characteristics of human interactions, such as the statistics of contact durations.
On the basis of this outcome, we provide a simple yet effective procedure to identify anomalous activity patterns in contact networks.

\spara{Polarized communities in signed networks.}
At first, we present a characterization of the polarized communities discovered by our spectral methods that proves the importance of the practical tweaks and the stability of the randomized solutions.
Then, to evaluate the proposed algorithms, we select non-trivial baselines inspired by methods proposed in literature for related problems.
Our algorithms produce higher quality solutions, are much faster than the baselines, and can scale to much larger networks.
In addition, they are able to identify ground-truth planted polarized communities in synthetic datasets.

A case study about Twitter data regarding the Italian Constitutional Referendum held on December 4, 2016~\cite{lai2018stance} provides tangible evidence for the goodness of our problem formulation and algorithms in identifying two communities that are polarized about a certain topic.
The solution returned by our algorithms almost perfectly split in two communities, favorable and against to the Referendum, the most active users included in the dataset, meaning that the solution identifies the ``core'' of the controversies, i.e., a set of intensely debating users about the Referendum.

\section{Conclusions and future work}
\label{sec:conclusions}
The problem of extracting cores and other dense structures from complex networks has emerged as a key graph-mining primitive in a variety of scenarios.
In this work, we propose novel definitions and methodologies for the identification of such structures in multilayer networks, temporal networks, and signed networks.
We show advancements in both theoretical aspects and practical applications.
In particular, we study ($i$) core decomposition in multilayer networks together with a series of its applications; ($ii$) core decomposition in temporal networks; and, ($iii$) the identification of polarized communities in a signed network.

This work opens several enticing avenues for further practical applications.
In our future investigation about multilayer networks, we plan to employ our core-decomposition tool for the analysis of multilayer brain networks to identify common patterns to patients affected by diseases or under the assumption of drugs and, also, to select features in order to discriminate actual patients from healthy individuals.
Similarly, we want to exploit span-cores features for network finger-printing and classification, model validation, and for new ways of visualizing large-scale time-varying graphs.
We will also study the role of maximal span-cores with large span in spreading processes on temporal (e.g., contact or presence) networks.
The application of the proposed definitions of polarized communities to real-world signed networks can have implications in computational social science problems, e.g., understanding opinion shifts in data streaming from social media sources.

Other future work will deal with more theoretical inquiries.
Specifically, we will exploit span-cores for the computation of related notions, such as \emph{community search} or \emph{densest subgraph} in temporal networks.
We also plan to study the tightness of our analysis of approximation of the two algorithms for the extraction of polarized communities from signed networks, and derive heuristic algorithms that return meaningful solutions in practice;
finally, it would be interesting to extend this problem to detect an arbitrary number of communities.




%


\end{document}